\documentclass[prl,showpacs,amsfonts,amssymb,amsmath,superscriptaddress,twocolumn]{revtex4}

\usepackage{graphicx}

\begin{document}

\title{Control and Tomography of a Three Level Superconducting Artificial Atom}

\author{R.~Bianchetti}
\author{S.~Filipp}
\author{M.~Baur}
\author{J.~M.~Fink}
\author{C.~Lang}
\author{L.~Steffen}
\affiliation{Department of Physics, ETH Zurich, CH-8093 Z{\"{u}}rich, Switzerland}
\author{M.~Boissonneault}
\author{A.~Blais}
\affiliation{D\'epartement de Physique, Universit\'e de Sherbrooke, J1K 2R1 Sherbrooke, Canada}
\author{A.~Wallraff}
\affiliation{Department of Physics, ETH Zurich, CH-8093 Z{\"{u}}rich, Switzerland}


\date{\today}

\begin{abstract}
A number of superconducting qubits, such as the transmon or the phase qubit, have an energy level structure with small anharmonicity. This allows for convenient access of higher excited states with similar frequencies. However, special care has to be taken to avoid unwanted higher-level populations when using short control pulses. Here we demonstrate the preparation of arbitrary three-level superposition states using optimal control techniques in a transmon. Performing dispersive read-out we extract the populations of all three levels of the qutrit and study the coherence of its excited states. Finally we demonstrate full quantum state tomography of the prepared qutrit states and evaluate the fidelities of a set of states, finding on average 96\%.
\end{abstract}

\maketitle

Spin 1/2 or equivalent two-level systems are the most common computational primitive for quantum information processing~\cite{Nielsen2000}. Using physical systems with higher dimensional Hilbert spaces instead of qubits has a number of potential advantages. They simplify quantum gates~\cite{Lanyon2009}, can naturally simulate physical systems with spin greater than 1/2~\cite{Neeley2009}, improve security in quantum key distribution~\cite{Cerf2002,Durt2004} and show stronger violations of local realism when prepared in entangled states~\cite{Kaszlikowski2000,Inoue2009}. Multilevel systems have been successfully realized in photon orbital angular momentum states~\cite{Mair2001,Molina-Terriza2004}, energy-time entangled qutrits~\cite{Thew2002} and polarization states of multiple photons~\cite{Vallone2007}. Multiple levels were used before for pump-probe readout of superconducting phase qubits~\cite{Martinis2002,Cooper2004,Lucero2008}, were observed in the nonlinear scaling of the Rabi frequency of DC SQUID's~\cite{Murali2004,Claudon2004,Dutta2008,Ferron2010} and were explicitly populated and used to emulate the dynamics of single spins~\cite{Neeley2009}. In solid state devices, the experimental demonstration of full quantum state tomography~\cite{Thew2002} of the generated states, i.e.~a full characterization of the qutrit, is currently actively pursued by a number of groups.

In this work, we use a transmon-type superconducting artificial atom with charging energy $E_{\rm{c}} / 2 \pi = 298\pm 1~{\rm MHz}$ and maximum Josephson energy $E_{\rm{J}}^{\rm{max}} / 2 \pi = 38$ GHz~\cite{Koch2007,Schreier2008} embedded in a coplanar microwave resonator of frequency $\omega_r/ 2 \pi = 6.9421\pm 0.0001~{\rm GHz}$ in an architecture known as circuit quantum electrodynamics (QED)~\cite{Blais2004,Wallraff2004b}. In circuit QED, the third level has already been used, for instance, in a measurement of the  Autler-Townes doublet in a pump-probe experiment~\cite{Baur2009,Sillanpaa2009}. It has also been crucial in the realization of the first quantum algorithms in superconducting circuits \cite{DiCarlo2009} and is used in a number of recent quantum optical investigations, e.g.~in Ref.~\cite{Abdumalikov2010}.  Also, quantum state tomography based on dispersive readout~\cite{Blais2004,Bianchetti2009} of a two-qubit system has been demonstrated~\cite{Filipp2009b} and used for the characterization of entangled states~\cite{DiCarlo2009,Leek2010}. In our realization of three level quantum state tomography, we populate excited states using optimal control techniques~\cite{Motzoi2009} and read out these states using tomography with high fidelity. We determine all relevant system parameters and compare the data to a quantitative model of the measurement response.

\begin{figure}[t!]
  \centering
    \centering
    \begin{minipage}{1.\linewidth}
        \includegraphics[width=1.\textwidth]{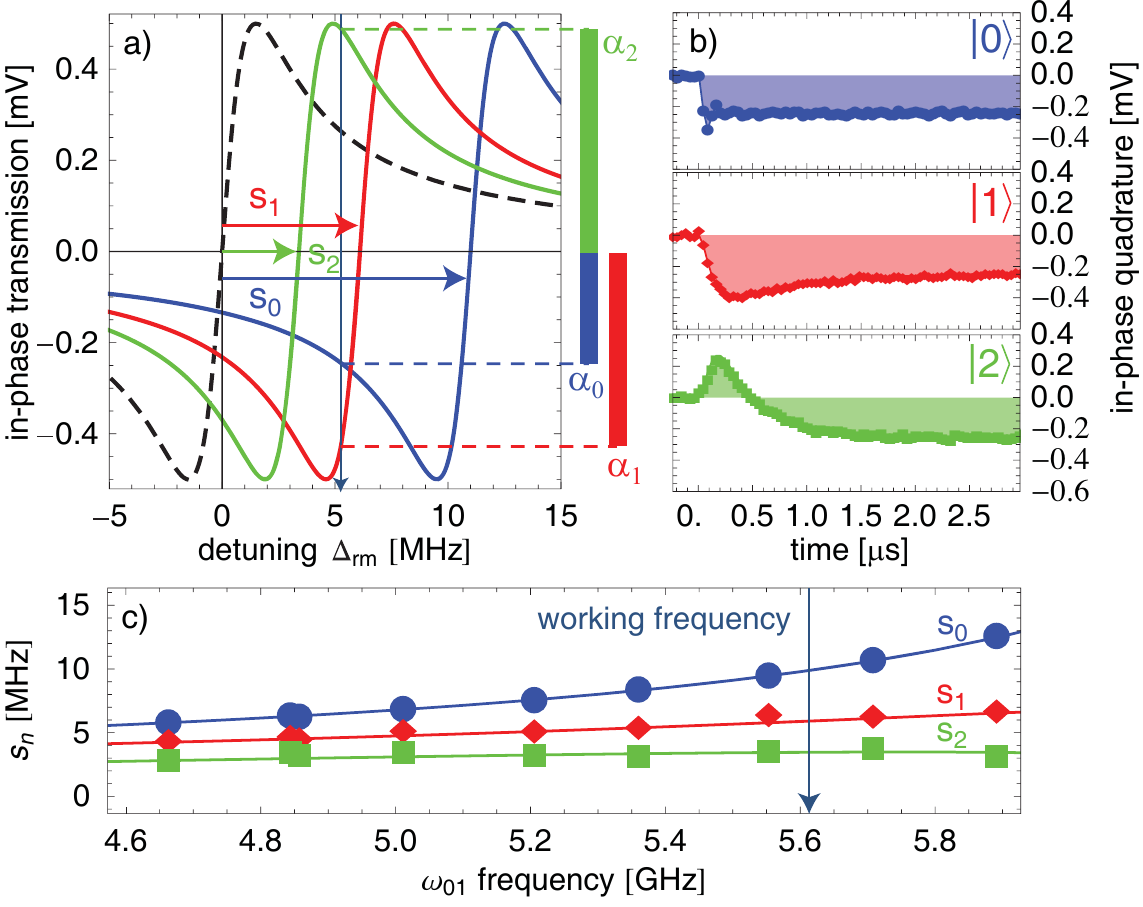}
    \end{minipage}
  \caption{(color online) a) Calculated in-phase transmission through the resonator for transmon states $n=0,1,2$. The dashed curve indicates the bare resonator response. The vertical blue arrow indicates the detuning $\Delta_{\rm{rm}}$ of the measurement tone. b) Pulsed I quadrature measurement responses for prepared states $|0\rangle$, $|1\rangle$ and $|2\rangle$. c) Measured dispersive shifts $s_n$ versus transmon transition frequency $\omega_{01}$. The solid lines are calculated within the linear dispersive approximation.}
    \label{fig:gef_levels}
\end{figure}

The transmon coupled to a single mode of a resonator is well described by the linear dispersive hamiltonian~\cite{Blais2004,Koch2007} approximated to second order
\begin{eqnarray}  \label{eq:dispersive_JC}
H_{JC}^D & = & \hbar \omega_r a^\dagger a + \sum_{n=0}^M \hbar \omega_n | n \rangle \langle n | + \sum_{n=1}^{M-1} \hbar \chi_{n-1} | n \rangle \langle n | \nonumber \\
 & & + \sum_{n=0}^{M-1} \hbar s_n | n \rangle \langle n | a^\dagger a,
\end{eqnarray}
where the transmon transition frequency $\omega_{01} = \omega_{1} - \omega_{0}$ is largely detuned from the resonator. Here, $a^{(\dagger)}$ is the annihilation (creation) operator for the photon field and $\chi_n = g^2_n/\Delta_n$. $g_n$ denotes the coupling strength to the transmon transition $n \leftrightarrow n+1$ and $\Delta_n$ the detuning of the same transition from the cavity frequency. We extracted $g_0 / 2 \pi = 115\pm 1~{\rm MHz}$ from a measurement of the vacuum Rabi mode splitting~\cite{Wallraff2004b}. Coupling constants $g_n$ of higher levels were explicitly determined in time resolved Rabi oscillation experiments, where $g_n = \eta_n g_0 \approx \sqrt{n+1} g_0$ due to the limited anharmonicity of the transmon ~\cite{Koch2007}. For the $\omega_{12}$ transition we experimentally determined $\eta_1 = 1.43 \pm 0.04$. Using flux bias, we detune the qubit by $\Delta_{0} = \omega_{01} -  \omega_r = -1.319 \pm 0.001~{\rm GHz}$ from the resonator. The non-resonant interaction with the transmon in state $|n\rangle$ leads to a dispersive shift $s_n = -(\chi_n - \chi_{n-1})$ in the cavity frequency. Measuring the in-phase quadrature amplitude~($\alpha^I_n$) of microwaves transmitted through the resonator [Fig.~\ref{fig:gef_levels}(a)] at a chosen detuning $\Delta_{\rm{rm}}= \omega_r-\omega_m =5.1~{\rm MHz}$ of the measurement frequency $\omega_m$ from $\omega_r$, allows to extract the population of the transmon state $n$.

To prepare arbitrary superposition states of the lowest three levels of the transmon we use optimal control techniques in which two subsequent DRAG pulses~\cite{Motzoi2009} of standard deviation 3~ns and total length 12~ns are applied to the qubit at the $\omega_{01}$ and $\omega_{12}$ transitions. We extend the technique described in~\cite{Motzoi2009} for the two lowest levels of the transmon to three levels using quadrature compensation and time-dependent phase ramps~\cite{Motzoi2010} to suppress population leakage to other states and to obtain well defined phases.

For a first characterization of the readout of higher levels, the transmon is prepared in one of its three lowest basis states $|i\rangle$ ($i=0,1,2$). After state preparation, a coherent microwave tone is applied to the cavity and the state dependent transmission amplitude is measured, Fig.~\ref{fig:gef_levels}(b). The amplitude of the tone was adjusted to maintain the average population of the cavity well below the critical photon number $n_{\rm{crit}}=\Delta_{0}^2/4 g_0^2=25$~\cite{Blais2004}. The time dependent transmission signals are characteristic for the prepared qubit states and agree well with the expected transmission calculated based on Cavity-Bloch equations \cite{Bianchetti2009}. We have generalized the formalism presented in Ref.~\onlinecite{Bianchetti2009} to three levels to quantitatively model the dispersive measurement. From the fits in Fig.~\ref{fig:gef_levels}(b), we have extracted the state dependent cavity frequency shifts $s_{0,1,2}/ 2 \pi = 10.0,5.9,3.4\pm 0.1~{\rm MHz}$, which are found to be within $0.1~{\rm MHz}$ of the values calculated from independently measured hamiltonian parameters. Also, the dispersive frequency shifts $s_n$ measured in this way agree well with the linear dispersive model over a wide range of transmon transition frequencies $\omega_{01}$, see Fig.~\ref{fig:gef_levels}(c).

\begin{figure}[t!]
  \centering
    \begin{minipage}{1.\linewidth}
        \includegraphics[width=1.\textwidth]{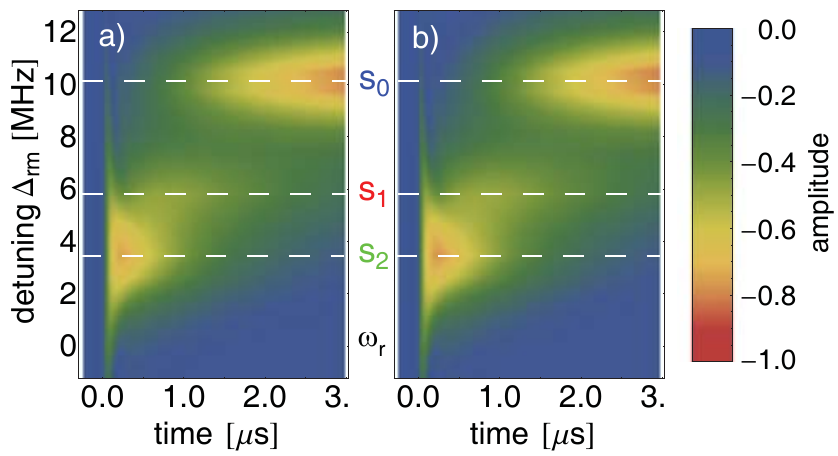}
    \end{minipage}
  \caption{(color online) a) Measured Q quadrature of the resonator transmission versus time and measurement detuning for  a preparation of the transmon in state $|2\rangle$. b) Calculation based on Cavity-Bloch equations.}
    \label{fig:fresponse}
\end{figure}

The frequency shifts can also be obtained more directly by measuring the transmission amplitude over a wide range of detunings $\Delta_{\rm{rm}}$ when preparing the transmon in the $|2\rangle$ state and observing its decay into the $|0\rangle$ state. Three distinct maxima in the measured $Q$ quadrature [Fig.~\ref{fig:fresponse}(a)] located at the expected frequencies shifted by an amount $s_n$ from $\omega_r$ are characteristic for the measurement of the $n=0,1,2$ states of the transmon. The peaks appear successively in time, as the transmon sequentially decays from $|2\rangle$ to $|1\rangle$ to the ground state $|0\rangle$. Sequential decay is expected due to the near harmonicity of the transmon qubit, for which only non-nearest-neighbor transitions are important~\cite{Koch2007}. The Q quadrature calculated from Cavity-Bloch equations is in good agreement with the measurement data and yields the energy relaxation times of the first and second excited state $T_1^{1}=800\pm 50~{\rm ns}$ and $T_1^{2}=700\pm 50~{\rm ns}$ as the only fit parameters, see Fig.~\ref{fig:fresponse}(b). The relaxation times are much longer than the typical time required to prepare the state using two consecutive 12~ns long pulses and allow for a maximum f-level population of $97 \%$, limited by population decay during state preparation~\cite{Chow2009}. The relative difference between data and calculated transmission is at most $3 \%$ at any given point indicating our ability to populate and measure the f-level with high fidelity.

To realize high-fidelity state tomography, arbitrary rotations in the Hilbert space with well defined phases and amplitudes are essential. Calibration of frequency, signal power and relative phases has to be performed based only on the population measurements of the qutrit states. To do so, we notice that the weak measurement partially projects the quantum state into one of its eigenstates $| 0 \rangle$, $| 1 \rangle$ or $| 2 \rangle$ in each preparation and measurement sequence~\cite{Blais2004, Filipp2009b, Bianchetti2009}. The average over many realizations of this sequence, which leads to the traces in Fig.~\ref{fig:gef_levels}(b), can therefore be described as a weighted sum over the contributions of the different measured states. This suggests the possibility of simultaneously  extracting the populations of all three levels from an averaged time-resolved measurement trace. Formally, the projective quantum non-demolition measurement gives rise to the following operator, which is diagonal in the three-level basis and linear in the population of the different states at all times,
\begin{equation} \label{eq:MQ}
\hat{M}_I(t)=\tilde{\alpha}_0(t) | 0 \rangle \langle 0 |+\tilde{\alpha}_1(t) | 1 \rangle \langle 1 |+\tilde{\alpha}_2(t) | 2 \rangle \langle 2 |.
\end{equation}
Here, $\tilde{\alpha}_n(t)$ are the averaged transmitted field amplitudes for the states $n$ sketched in Fig~\ref{fig:gef_levels}(a). The transmitted in-phase quadrature
\begin{equation} \label{eq:Q}
\langle I_{\rho}(t) \rangle =  {\rm Tr}\left[ \hat{\rho} \hat{M}_I(t) \right] = p_0 \tilde{\alpha}_0(t)+ p_1 \tilde{\alpha}_1(t) +p_2 \tilde{\alpha}_2(t),
\end{equation}
can be calculated for an arbitrary input state with density matrix $\rho$ and populations $p_i$. Since any measured response is a linear combination of the known pure $| 0 \rangle$, $| 1 \rangle$ and $| 2 \rangle$ state responses weighted by $p_i$,  the populations can be reconstructed using an ordinary least squares linear regression analysis, which pseudo-inverts Eq.~(\ref{eq:Q}) for each time step $t_i$. The reconstructed populations show larger statistical fluctuations than in the two-level case~\cite{Bianchetti2009} due to the pseudo-inversion of the ill-conditioned matrix used to calculate the $p_i$ from Eq.~(\ref{eq:Q}). The statistical error is influenced by the distinguishability between the different traces, see Fig.~\ref{fig:gef_levels}(b), and is minimized by optimizing the measurement detuning. In contrast to full quantum state tomography (see below), this method does not require any additional pulses after the state preparation. It is therefore used to find the $\omega_{12}$ transition frequency and the pulse amplitudes needed to generate accurate pulses for tomography.
\begin{figure}[t!]
  \centering
    \begin{minipage}{0.9\linewidth}
        \includegraphics[width=1.\textwidth]{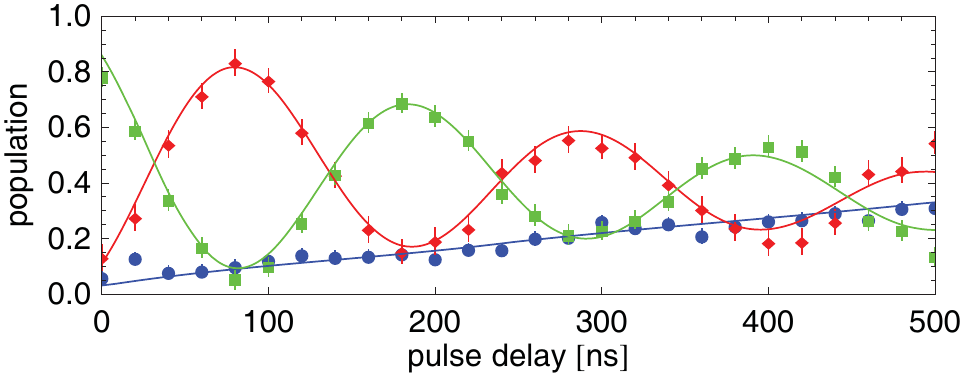}
    \end{minipage}
  \caption{(color online) Reconstructed transmon populations in a Ramsey oscillations experiment on the 1-2 transition using a 5~MHz detuned drive field. For a given pulse delay $| 0 \rangle$ (blue dots), $| 1 \rangle$ (red diamonds) and $| 2 \rangle$ (green squares) populations are extracted from a time resolved averaged measurement trace. The lines are calculated using Bloch equations.}
    \label{fig:Sparameters_efRamsey}
\end{figure}

The pulse amplitudes are extracted from Rabi-oscillations. To asses the precise value of $\omega_{12}$, we perform a Ramsey experiment between the $| 1 \rangle$ and the $| 2 \rangle$ level, see Fig.~\ref{fig:Sparameters_efRamsey}. We apply a $\pi$-pulse at $\omega_{01}$ and then delay the time between two successive $\pi/2$ pulses applied at $\omega_{12}$ before starting the measurement. The theoretical lines are calculated based on a Bloch equation simulation with a dephasing time of the 2-level fitted to $T_2^2=500~{\rm ns}$.

Using quantum state tomography~\cite{Thew2002}, the full density matrix of the first three levels of a transmon can be reconstructed. This is achieved by performing a complete set of nine independent measurements after preparation of a given state and calculating the density matrix based on the measurements outcomes. Since the measurement basis is fixed by our hamiltonian Eq.~(\ref{eq:dispersive_JC}), the state is rotated by applying the following pulses prior to measurement: $\mathbb{I}$, $\left(\frac{\pi}{2}\right)_x^{01}$,\,$\left(\frac{\pi}{2}\right)_y^{01}$,\,$\left(\pi\right)_x^{01}$,\,$\left(\frac{\pi}{2}\right)_x^{12}$,\,$\left(\frac{\pi}{2}\right)_y^{12}$, $\left(\pi\right)_x^{01}\left(\frac{\pi}{2}\right)_x^{12}$,\,$\left(\pi\right)_x^{01}\left(\frac{\pi}{2}\right)_y^{12}$,\,$\left(\pi\right)_x^{01}\left(\pi\right)_x^{12}$, where $\mathbb{I}$ denotes the identity and $\left(\theta\right)_a^{ij}$ denotes a pulse of angle $\theta$ on the $ij$-transition about the $a$-axis. For each of these unitary rotations $\left( U_k \right)$ we measure the coefficients $\langle I_k \rangle \equiv \text{Tr}[\rho U_k \hat{M}_I U_k^\dagger]$ by integrating the transmitted in-phase quadrature in Eq.~(\ref{eq:MQ}) over the measurement time~\cite{Filipp2009b}, i.e.~implementing the measurement operator $\hat{M}_I = \int_0^T \hat{M}_I(t) \, dt$. This relation is inverted to reconstruct the density matrix $\rho$ by inserting the known operators $U_k \hat{M}_I U_k^\dagger$. Note, that unlike in the preceding measurement of the populations only, we now extract a single quantity, $\langle I_k\rangle$, for each measured time trace. Quantum state tomography based on the simultaneous extraction of the populations of $| 0 \rangle$, $| 1 \rangle$ and $| 2 \rangle$ could potentially reduce the number of required measurements, but might come at the expense of larger statistical errors as discussed above.

\begin{figure}[t!]
  \centering
    \begin{minipage}{1.\linewidth}
        \includegraphics[width=1.\textwidth]{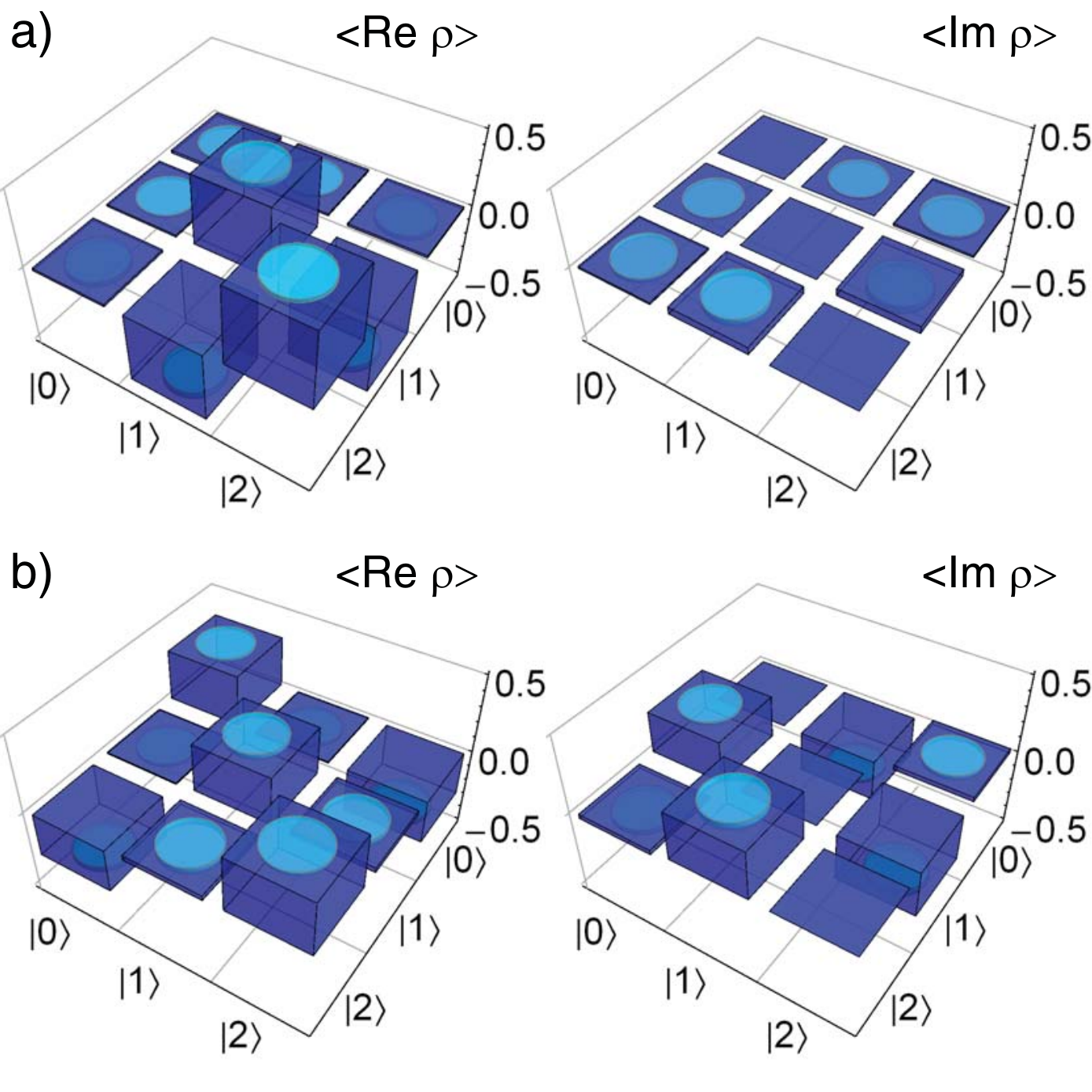}
    \end{minipage}
  \caption{(color online) Measured Real and imaginary part of the reconstructed density matrices of $|\Psi_a\rangle = 1/\sqrt{2} \left( | 1 \rangle - | 2 \rangle \right)$ and $|\Psi_b\rangle = 1/\sqrt{3} \left( | 0 \rangle + i | 1 \rangle - | 2 \rangle \right)$. The cyan cylinders indicate the standard deviations, typically 0.02.}
    \label{fig:tom}
\end{figure}

Examples of measured density matrices are shown in Fig.~\ref{fig:tom} for the states $|\Psi_a\rangle = 1/\sqrt{2} \left( | 1 \rangle - | 2 \rangle \right)$ and $|\Psi_b\rangle = 1/\sqrt{3} \left( | 0 \rangle + i | 1 \rangle - | 2 \rangle \right)$. A maximum likelihood estimation procedure has been implemented~\cite{James2001}. The extracted fidelities $\emph{F} \equiv \langle \psi | \rho | \psi \rangle$ of $97\pm 2\%$ and $92\pm 2\%$, respectively, demonstrate the high level of control and the good understanding of the readout of our three level system. Considering the measured decay rates, the best achievable fidelity for the states $|\Psi_{a/b}\rangle$ is $97\pm 1\%$. Preparing a set of 14 different states we measure an average fidelity of $96 \%$, with a minimum of $92\pm 2\%$ for the pure  $| 2 \rangle$ state. The small remaining imperfections are likely due to phase errors in the DRAG pulses which affect both state preparation and tomography.

We have demonstrated the preparation and tomographic reconstruction of arbitrary three level states in a superconducting quantum circuit. Controlling and reading out higher excited states in these systems does broaden the prospects of using such circuits for future experiments in the domains of quantum information science and quantum optics.

We thank Marek Pikulski for his contributions to the project. This work was supported by SNF, the EU projects SOLID and EuroSQIP and ETH Zurich. AB is supported by NSERC, the Alfred P. Sloan Foundation and is a CIFAR Scholar


\end{document}